\documentclass[12pt,a4paper]{article}
\usepackage{graphicx,epsfig}
\usepackage{dcolumn} 
\usepackage{slashed,color,amsmath,amssymb}
\usepackage{jheppub}
\usepackage{enumitem}
\usepackage{siunitx}
\usepackage{bm}
\usepackage{layouts}
\usepackage[svgnames]{xcolor}
\usepackage{float}
\usepackage{caption, subcaption}
\usepackage{multirow}
\usepackage{verbatim}
\usepackage[normalem]{ulem}

\usepackage[compat = 1.1.0]{tikz-feynman}
\usepackage{tikz}
\newcommand{\be}{\begin{equation}}
\newcommand{\ee}{\end{equation}}
\newcommand{\bea}{\begin{eqnarray}}
\newcommand{\eea}{\end{eqnarray}}
\newcommand{\bit}{\begin{itemize}}
\newcommand{\eit}{\end{itemize}}

\newcommand{\lam}{\lambda}

\definecolor{greeen}{HTML}{008ae6}

\definecolor{newgreen}{HTML}{009900}
\definecolor{newnewgreen}{HTML}{21db2c}

\newcommand\thefontsize{The current font size is: \f@size pt}

  \definecolor{fsred}{HTML}{c71616}

\definecolor{fsblue}{HTML}{1e88e5}

\definecolor{fsyellow}{HTML}{dca607}

\definecolor{fsgreen}{HTML}{014d40}

\definecolor{fsviolet}{HTML}{a11b9f}

\definecolor{newpurple}{HTML}{7B1DAA}

\def\gsim{\lower0.5ex\hbox{$\:\buildrel >\over\sim\:$}}
\def\lsim{\lower0.5ex\hbox{$\:\buildrel <\over\sim\:$}}

\hypersetup{
   colorlinks=true,       
   linkcolor=blue,        
   citecolor=red,         
   filecolor=magenta,     
}

\DeclareRobustCommand{\orderof}{\ensuremath{\mathcal{O}}}
\def\neutralino{{\ensuremath{\tilde{\chi}^0_1}}}
\def\chiobs{\ensuremath{N^{\mathrm{obs}}_{\neutralino{}}}}

\def\ATLAS{\texttt{ATLAS}}
\def\CMS{\texttt{CMS}}

\def\ANUBIS{\texttt{ANUBIS}}
\def\CODEXB{\texttt{CODEX-b}}
\def\FACET{\texttt{FACET}}
\def\FASER{\texttt{FASER}}
\def\FASERTwo{\texttt{FASER2}}
\def\MAPP{\texttt{MAPP}}
\def\MAPPTwo{\texttt{MAPP2}}
\def\MATHUSLA{\texttt{MATHUSLA}}
\def\FORESEE{\texttt{FORESEE}}
\def\PYTHIA{\texttt{Pythia 8.313}}

\def\stilde{\widetilde}

\preprint{\begin{flushright} BONN-TH-2026-14
	\end{flushright}}	

\graphicspath{{./figs/}}

\title{The Single Photon Signature of a Light Long-lived
Neutralino at Remote Detectors at the LHC}

    \author[a]{Herbi\,K.\,Dreiner,}
	\emailAdd{dreiner@uni-bonn.de}
	\affiliation[a]{Bethe Center for Theoretical Physics \& Physikalisches Institut der Universit\"at Bonn,\\ Nu{\ss}allee 12, 53115 Bonn, Germany}
    \author[a]{Julian\,G\"unther,}
	\emailAdd{jyahguen@uni-bonn.de}
	
	\author[a]{Dominik\,K\"ohler,}
	\emailAdd{dkoehler@uni-bonn.de}
	
	\author[a]{Apoorva\,Shah}
	\emailAdd{ashah@uni-bonn.de}

\abstract{We investigate the phenomenology of light 
long-lived neutralinos in \mbox{R-parity} violating supersymmetric 
models, focusing on the proposed remote detectors
\ANUBIS{}, \CODEXB{}, \FACET{}, \FASER{}, \FASERTwo{}, 
\MAPP{}, \MAPPTwo{}, and \MATHUSLA{} at the LHC. We assume
the production of the neutralinos at the \texttt{ATLAS} or 
\texttt{CMS} interaction points via rare scalar 
meson decays induced by R-parity violating couplings. We 
study six supersymmetric R-parity violating benchmark 
scenarios in which the dominant neutralino decay is $
\neutralino{}\to\gamma+\nu$. For each scenario, we determine the
projected search sensitivity at the above listed detectors. 
Extending previous work focused primarily on \FASER{} and 
\FASERTwo{}, we improve the simulation by taking into 
account the extended flight path of the parent meson. 
We find that \ANUBIS{} provides the best sensitivity to our 
benchmark scenarios and \FASER{} the least among the 
considered experiments, while of course \FASER{} has already 
taken data.}

\begin{document}
\maketitle



\section{Introduction}
\label{sec:intro}

Supersymmetry (SUSY) is a theoretically well-motivated extension 
of the Standard Model of particle physics (SM) 
\cite{WessZumino1974a, WessZumino1974b, Haag1975}. In its minimal 
form, the so-called Minimal Supersymmetric Standard Model (MSSM) 
provides a good framework to address many of the open questions
of the SM, such as the neutrino masses, the baryon asymmetry and
the dark matter in the Universe. For reviews, see for example 
Refs.~\cite{Nilles:1983ge, HaberKane1985, Martin:1997ns, 
Baer:2006rs, twocomponenttextbook}. As a promising solution to the 
hierarchy problem \cite{Veltman:1980mj}, supersymmetry could very 
well manifest itself at energies within reach of current 
experiments. Efforts searching for physics beyond the SM are 
underway, spanning a wide range of experimental and observational 
approaches. These include collider experiments~\cite{Lee2019} with 
near or far detectors \cite{Alimena:2019zri}, beam-dump 
studies~\cite{Cesarotti2023,Jodlowski2020}, nuclear and electron 
recoil measurements~\cite{Aalbers2023}, as well as astrophysical 
observations~\cite{Planck2018, SuperK2016, IceCube2017}.

The \ATLAS{}~\cite{ATLAS2008,ATLAS:2023dns} and \CMS{}~\cite{CMS2008, 
CMS:2023gfb} collaborations at the LHC have been devoting a large
effort in the search for SUSY. See for example
Refs.~\cite{ATLAS:2020syg, CMS:2021beq}. See also the overview of 
supersymmetry searches compiled by the Particle Data Group 
\cite{ParticleDataGroup:2024cfk}. Although no definitive discovery 
of such particles has yet been achieved, stringent lower bounds on 
their masses have been established. For example, squark and gluino 
masses are constrained to be higher than $\mathcal{O}(5\,$TeV), 
depending on the precise supersymmetric model~\cite{ATLAS:2024vyj,
CMS:2025bxo}.

While most collider searches have focused on R-parity conserving 
(RPC) SUSY, which typically involves missing transverse momentum 
signatures, models with R-parity violation (RPV) are equally 
well-motivated \cite{Ibanez:1991pr, Dreiner:2005rd, Dreiner:2006xw,
Dreiner:2012ae}. In addition, light neutrinos can arise naturally 
in RPV models without introducing a new see-saw energy scale 
\cite{Minkowski:1977sc, Dreiner:2007yz, Dreiner:2007vp, 
Dreiner:2006xw}. Note, in RPV scenarios, any superpartner can in 
principle be the lightest supersymmetric particle (LSP), as there 
is no dark matter constraint \cite{Dreiner:2008ca, Desch:2010gi, 
Dercks:2017lfq}. Furthermore, the LSP can possibly decay within the 
detector. Therefore, RPV scenarios entail a different and richer 
phenomenology of signatures \cite{Hall:1983id, Dreiner:1991pe}
than RPC. At this point, we note Wagner's conjecture on signature 
anomalies at the LHC \cite{Bittar:2025rcw}. If it holds, it would 
state that a comprehensive coverage of new physics is possible with
RPV searches, see also Refs.~\cite{Dreiner:1991pe, Dercks:2017lfq, 
Dreiner:2023bvs, Dreiner:2025kfd}. We take this as a further 
motivation to investigate the broad range of RPV signatures in 
existing and planned experiments. Here, we focus specifically on 
signatures resulting from a light long-lived neutralino LSP, 
$\neutralino{}$.

Within the RPV scenarios, the lightest neutralino is the LSP in 
large regions of the RPV-MSSM parameter space 
\cite{Allanach:2003eb, Dreiner:2008ca, Dercks:2017lfq, 
Hempfling:1995wj, Allanach:1999mh}. Going beyond the RPV-MSSM and 
allowing for non-universal gaugino masses, the neutralino is still 
the LSP in large regions of parameter space, furthermore, it can be 
very light \cite{Choudhury:1999tn, Dedes:2001zia, Gogoladze:2002xp}.
In fact, a massless neutralino is consistent with all laboratory and 
observational data \cite{Dreiner:2003wh, Dreiner:2009ic, 
Dreiner:2011fp}. We emphasize, a light neutralino in the MeV mass 
range cannot constitute the dark matter of the Universe, as it 
violates the Lee-Weinberg bound \cite{Lee:1977ua, Hooper:2002nq, 
Belanger:2002nr, Calibbi:2014lga, Barman:2022jdg, Barman:2024xlc},
it must decay.

A neutralino with mass lighter than a few GeV is necessarily 
dominantly bino-like \cite{Choudhury:1999tn, Dreiner:2009ic}. Thus, 
at tree-level, it does not couple to the $Z^0$ boson and contributes 
only minimally to its invisible width \cite{Dreiner:2009ic}. Due to its
small mass, such a light neutralino can be long-lived on collider 
experiment time scales, while decaying via RPV-induced 
interactions~\cite{Dreiner:1997uz,twocomponenttextbook,Barbier:2004ez}.
Such long-lived states are particularly interesting since their 
displaced or delayed signatures are less constrained by existing 
collider searches and may be more effectively probed at dedicated 
long-lived particle (LLP) experiments. We provide more details in 
Section~\ref{sec:theory}. For a neutralino with mass $m_{\neutralino{}}
\lsim\mathcal{O}(1\,\mathrm{GeV})$, the leading possible decay modes at 
tree-level are \cite{Domingo:2022emr}
\begin{eqnarray}
\neutralino{}\to \left\{
\begin{array}{lc}
  \ell^+_i\ell^{\prime -}_k\nu_j,\, \nu_i\nu_j\nu_k,\;   &  \mathrm{L_iL_j\bar 
  E_k}\,,\\[2mm]
 \ell_i^\pm M_{jk}^\mp,\, \nu_i M_{jk}^0,\;   & \mathrm{L_iQ_j\bar D_k}\,, \\[2mm]
p+ M^-,\, n+ M^{\prime 0},\;&  \mathrm{\bar U_1 \bar D_1 \bar D_2}\,.
\end{array}
\right.
\end{eqnarray}
Here $\ell^\pm_{i,k}\in\{e^\pm,\;\mu^\pm\}$ are light charged 
leptons, the $\nu_{i,j,k}$ denote the 3 possible neutrinos, and we
write $M^{(0,\pm,')}$ for the various potential meson final 
states. In all cases the charge conjugate final state is also 
allowed, as the $\neutralino{}$ is a Majorana fermion. On the right, 
we list the dominant RPV operator of the decay with generation indices.

In this paper, we focus on an interesting signature, which only 
arises at 1-loop order in RPV-SUSY. The neutralino can decay to just
a neutrino and a photon via operators of the form $L_iL_j\bar E_j$ 
or $L_iQ_j\bar D_j$ \cite{Hall:1983id, Dawson:1985vr}
\begin{align}
    \neutralino{} \rightarrow \gamma + \overset{(-)}{\nu_i}.
    \label{eq:gamma-nu}
\end{align}
This signature is also well-motivated by theories such as 
gauge-mediated supersymmetry breaking (GMSB)~\cite{Ambrosanio1997}, 
extensions involving  hidden or dark sectors that couple feebly to 
the SM~\cite{Essig2013}, and  models with universal extra 
dimensions~\cite{Appelquist2001}. In all these cases, the 
combination of high-energy photons with missing transverse energy 
provides a clean and experimentally robust probe of BSM dynamics, 
making it a key search channel at current and future collider 
experiments. This opens a new window for discovery.

Such final states have been extensively studied in context of 
\ATLAS{} and \CMS{}~\cite{ATLAS:2014kci,ATLAS:2018nud,
CMS:2012bbi, CMS:2017brl,CMS:2019vzo,CMS:2018ffd}. 
Ref.~\cite{Dreiner:2022swd} investigates the capabilities of the 
\FASER{} and \FASERTwo{}~\cite{FASER:2018bac, 
FASER2:DesignPerformance, FASER2019} detectors to look for such a 
signal using the simulation tool \FORESEE{} within RPV-SUSY. 
However, many existing and proposed collider-based experiments are capable of probing 
such a signal, including  \FACET{}~\cite{Cerci:2021nlb}, 
\MATHUSLA{}~\cite{MATHUSLA2019}, 
\CODEXB{}~\cite{Gligorov:2017nwh,CODEX-b:2024tdl}, 
\ANUBIS{}~\cite{Bauer:2019vqk}, and 
\texttt{MoEDAL-MAPP}~\cite{MoEDAL-MAPP:2022kyr}, which is in two 
phases: \MAPP{} and \MAPPTwo{}. Here we go 
beyond the work of Ref.~\cite{Dreiner:2022swd} to investigate the signature Eq.~\eqref{eq:gamma-nu} at the additional collider-based experiments listed 
above. We also go beyond the simulation technique of 
Ref.~\cite{Dreiner:2022swd}, which we describe in detail in 
Section~\ref{sec:simulation}. We thus investigate this search channel 
for a broad set of experiments using a different simulation method, and
perform a comparative sensitivity study of these experiments under well
motivated benchmark scenarios. Note, we restrict ourselves here to 
collider-based experiments. We leave a study of the single-photon signature 
at fixed-target experiments such as \texttt{DUNE}~\cite{DUNE:TDRVol1, DUNE:TDRVol2, 
DUNE:TDRVol3, DUNE:TDRVol4} and \texttt{SHiP} \cite{deVries:2015mfw, SHiP:2018xqw, SHiP:2021nfo},
or at neutrino reactor experiments such as \texttt{JUNO}~\cite{JUNO:2021vlw} for a later study.

This paper is organised as follows. Section~\ref{sec:theory} provides 
the theoretical framework of the RPV-MSSM and the lightest neutralino 
in detail. The simulation strategy for the signal estimation is 
described in Section~\ref{sec:simulation}. The outcomes of the 
simulations performed for various experimental setups are presented in 
Section~\ref{sec:results}. Finally, Section~\ref{sec:conclusion} 
provides a summary of the results and concluding remarks.

\section{Theoretical Framework}
\label{sec:theory}
\subsection{The RPV-MSSM}
Given the particle content of the MSSM and the structure of the $N=1$
supersymmetry algebra, the most general renormalizable and $SU(3)_C 
\times SU(2)_L \times U(1)_Y$-invariant superpotential can be written
as~\cite{Allanach:2003eb}
\begin{align}
    W = W_{\text{MSSM}} + W_{\text{LNV}} + W_{\text{BNV}},
\end{align}
where $W_{\text{MSSM}}$ is the usual MSSM 
superpotential, and, the RPV terms are given 
by
\begin{align}
W_{\text{LNV}} =\frac{1}{2} \, \lambda_{ijk} \, L_i L_j \bar{E}_k
+ \lambda'_{ijk} \, L_i Q_j \bar{D}_k + \kappa_i H_u L_i \,,\quad W_{\text{BNV}}=\frac{1}{2} \, \lambda''_{ijk} \, \bar{U}_i \bar{D}_j \bar{D}_k\,.
\end{align}
The additional terms $W_{\text{LNV}}$ and $W_{\text{BNV}}$ introduce 
lepton-number-violating (LNV) and baryon-number-violating (BNV) 
interactions, respectively. $L_i (Q_i)$ and $ \bar{E}_i (\bar{U}_i, 
\bar{D}_i)$ denote the lepton (quark) $SU(2)_L$ doublets and $SU(2)_L$ 
singlet chiral superfields, respectively. $H_u$ is the up-type Higgs 
doublet. The indices $i, j,k \in\{ 1,2,3\}$ represent generation indices
with a summation implied over repeated indices. The $\lambda, \lambda',$
and $\lambda''$ are dimensionless coupling constants. The $\kappa_i$
are dimension-one mass mixing parameters. We shall set them to zero
in our analysis. This corresponds to a basis choice 
\cite{Hempfling:1995wj, Hirsch:2000ef, Allanach:2003eb, deCampos:2007bn,
HallSuzuki1984, DreinerThormeier2004}. We have suppressed gauge indices 
here.

While $W_{\text{LNV}}$ and $W_{\text{BNV}}$ are usually set to zero by 
imposing RPC to ensure proton stability, some RPV interactions may still
be allowed if only specific terms are present~\cite{Barbier:2004ez, 
DreinerThormeier2004,Chemtob:2004xr,Chamoun2021, PDG2024}, or if the 
associated couplings are sufficiently small to satisfy experimental 
constraints~\cite{Domingo:2024qoj, Dreiner:2020qbi, Wang:2023trd, 
Dreiner:2010ye, Dreiner:2023bvs,Dreiner:2025kfd}. In the following we 
shall consider non-zero values for the $\lambda_{ijk},\,\lambda'_{ijk}$
LNV couplings.

\subsection{A Long-lived Light Neutralino}
In RPV-SUSY, any SUSY particle can be the LSP~\cite{Dercks:2017lfq,
Dreiner:2008ca, Dreiner:2009fi, Desch:2010gi}. We shall consider the 
case of a neutralino LSP here. Neutrinos can be produced indirectly via 
the strong interaction via meson decays, \textit{e.g.} pions: $\pi^+\to 
\mu^++\nu_\mu$. Similarly neutralinos can also be singly produced, for 
example, via the decays of pseudoscalar mesons. This involves the
$LQ\bar{D}$ operators with $\lam'$ couplings. The corresponding decay
widths are provided in Ref.~\cite{deVries:2015mfw}, which we show here 
using the same conventions:
\begin{equation}
    \Gamma\left(M_{a b} \rightarrow 
    \neutralino{}+l_i\right)=\frac{\lambda^{\frac{1}{2}}\left(m_{M_{a 
    b}}^2, m_{\neutralino{}}^2, m_{l_i}^2\right)}{64 \pi m_{M_{a 
    b}}^3}\left|G_{i a b}^{S, f}\right|^2\left(f_{M_{a 
    b}}^S\right)^2\left(m_{M_{a 
    b}}^2-m_{\neutralino{}}^2-m_{l_i}^2\right).
    \label{eq:rpv-meson-decay}
\end{equation} 
Here, $l_i$ represents either a charged lepton $\ell_i^{\pm}$ or a 
neutrino $\nu_i$, depending on whether the meson $M_{ab}$ is charged or 
neutral. The $G_{i a b}^{S, f}$ are given in terms of linear 
combinations of various $\lam'/m_{\text{SUSY}}^2$, where $\lam'$ is the 
RPV coupling involved in the production of the neutralino and $m_{\text 
{SUSY}}$ is the mass of the SUSY particle mediating the decay  
\cite{deVries:2015mfw}. $a,b$ are the generation indices of the 
quark/antiquark bound in the meson. $m_{M_{ab}},\, m_{\neutralino{}},$ 
and $m_{l_i}$ denote the meson, the neutralino and the lepton mass, 
respectively. $\lam ^{\frac{1}{2}}
(x,y,z)\equiv \sqrt{x^2+y^2+z^2-2xy-2yz-2xz}$ is the square root of the K\"{a}ll\'{e}n function. The $f_{M_{ab}}^S$ denote the pseudoscalar meson decay constants.
 
As mentioned, the lightest neutralino $\neutralino{}$ can be very light, 
even massless~\cite{Dreiner:2011fp}. Such a light neutralino with mass 
below about 1~GeV can decay via RPV interactions leading to so-called 
“displaced vertices”, if the RPV couplings are sufficiently small. The 
decay vertex can be displaced by meters or even 100s of meters 
\cite{Choudhury:1999tn, Dedes:2001zia, Dercks:2018eua, Dreiner:2022swd, 
Gunther:2023vmz}. Several dedicated far-detector programs have been 
proposed near the LHC interaction points (IPs), primarily designed to 
search for such long-lived particles (LLPs) with just such decay lengths 
on the order of $c\tau \sim(1\!-\!100)\,\mathrm{m}$, or longer. 

At loop level, the neutralino can decay radiatively 
to a photon and a neutrino through the $W_{\text{LNV}}$ terms, as discussed in~\cite{HallSuzuki1984, 
Dawson:1985vr, Dercks:2018eua, Dreiner:2022swd}, \textit{cf.} 
Eq.~\eqref{eq:gamma-nu}. Since the final state consists of a photon 
accompanied by missing energy carried away by the neutrino, this decay 
channel offers a particularly clean signature with low backgrounds in 
experiments. In addition, it enables sensitivity to lower-mass neutralino
scenarios that might otherwise be inaccessible, as the final state has 
effectively no mass threshold. The decay width of the process is 
\begin{equation}
    \Gamma(\neutralino{} \rightarrow \gamma \overset{(-)}{\nu_i}) = \frac{\alpha^2 \lam^{(\prime)2} m^3_{\neutralino{}}}{512 \pi^3\cos{\theta_W}} \left[\sum_f e_f N_c m_f\frac{(4e_f + 1)}{m^2_{\tilde{f}}} \left( 1+ \log\left(\frac{m^2_{f}}{m^2_{\tilde{f}}}\right)\right)\right]^2,
\label{eq:neutralinodecay}
\end{equation}
where $\lambda^{(')}$ denotes the relevant RPV coupling which contributes
to the process. Note, that these have to have the index structure: $\lam_{ijj}$
or $\lam'_{ijj}$. The notation in Eq.~\eqref{eq:neutralinodecay} is the 
same as Ref.~\cite{Dreiner:2022swd}. $e_f$, $m_f (m_{\tilde{f}})$ and 
$C_f$ are the electric charge, mass and the color factors ($C_f=3$ for 
quarks, $C_f=1$ for leptons) of the fermions (sfermions) in the loop, 
respectively. $\alpha$ is the fine structure constant, $\theta_W$ is the 
electroweak mixing angle. For the $\lam$ ($LL\bar{E}$) coupling, the loop 
contains a $\ell_i-\tilde{\ell}_i$ pair, where $\ell$ is a charged 
lepton. For the $\lam'$ ($LQ\bar{D}$) coupling, a $d_i-\tilde{d}_i$ pair 
is required. In the decay width, the $b$ quark gives the largest 
contribution in the $\lam'$ case due to its large mass, note the $\sim
m^2_f$ dependence. For the $\lam$ case, the $\tau$ in the loop gives the 
highest contribution. A summary of upper bounds on the relevant RPV 
couplings used in this study is provided in
Tab.~\ref{tab:current_rpv_constraints}, which are obtained from 
Ref.~\cite{Allanach:1999ic, Dercks:2017lfq}.

\begin{table}[ht!]
	\centering
	\begin{tabular}{|c|cc|}
		\hline
		$ijk$ & $\lambda_{ijk}$															&	$\lambda^\prime_{ijk}$	\\
		\hline
		112	&	$-$																		&	$0.21\big(\frac{m_{\stilde s_R}}{\SI{1}{\tera\eV}}\big)$	\\[3mm]
		211	&	$0.49\big(\frac{m_{\stilde e_R}}{\SI{1}{\tera\eV}}\big)$				&	$0.59\big(\frac{m_{\stilde d_R}}{\SI{1}{\tera\eV}}\big)$	\\[3mm]
		212	&	$0.49\big(\frac{m_{\stilde \mu_R}}{\SI{1}{\tera\eV}}\big)$				&	$0.59\big(\frac{m_{\stilde s_R}}{\SI{1}{\tera\eV}}\big)$	\\
		221	&	$-$																		&	1.12	\\
		222	&	$-$																		&	1.12	\\
		313	&	$0.019\big(\frac{m_{\stilde \tau}}{\SI{1}{\tera\eV}}\big)^\frac{1}{2}$	&	1.12	\\[3mm]
		322	&	$0.70\big(\frac{m_{\stilde \mu_R}}{\SI{1}{\tera\eV}}\big)$				&	1.12	\\
		333	&	$-$																		&	1.04	\\
		\hline
	\end{tabular}
	\caption{Summary of current constraints of the $\lambda_{ijk}$ ($L_iL_j\bar{E}_k$) and $\lam^\prime_{ijk}$ ($L_iQ_j\bar{D}_k$) couplings \cite{Dercks:2017lfq,Allanach:1999ic}. The selected generation indices $ijk$ are present in the benchmarks investigated in Section~\ref{sec:results}.}
	\label{tab:current_rpv_constraints}
\end{table}

\section{Simulation Procedure}
\label{sec:simulation}

\begin{table}
	\centering
	\begin{tabular}{|c|cc|cccccc|}
		\hline
		Meson $M$ 	& $\pi^\pm$ 			& $\pi^0$ 				& \multicolumn{2}{c}{$K^\pm$} 				& \multicolumn{2}{c}{$K_L$} 				& \multicolumn{2}{c|}{$K_S$} \\
		$N_M$ 		& $1.64\times 10^{19}$ 	& $9.24\times 10^{18}$ 	& \multicolumn{2}{c}{$2.38\times 10^{18}$} 	& \multicolumn{2}{c}{$1.30\times 10^{18}$} 	& \multicolumn{2}{c|}{$1.31\times 10^{18}$} \\
		\hline
		\hline
		Meson $M$ 	& $D^\pm$ 				& $D^\pm_s$  		& & \multicolumn{2}{c}{$B^\pm$} 				& \multicolumn{3}{c|}{$B^0/\bar{B}^0$} \\
		$N_M$ 		& $2.03\times 10^{16}$ 	& $6.62\times 10^{15}$ 	& & \multicolumn{2}{c}{$1.46\times 10^{15}$} 	& \multicolumn{3}{c|}{$1.46\times 10^{15}$}  \\
		\hline				
	\end{tabular}
	\caption{Number of mesons over the total $4\pi$ solid angle produced in $pp$ collisions at the LHC with a center-of-mass energy of \SI{14}{\tera\eV} and an integrated luminosity of \SI{3}{\per\atto\barn}. For Kaons, $D$- and $B$-mesons we follow Ref.~\cite{DeVries:2020jbs,Gunther:2023vmz}, while for pions we obtain $N_M$ with the cross sections provided by \FORESEE{} \cite{Kling:2021fwx}.\protect\footnotemark}\label{tab:meson_numbers}
\end{table}

Within the light neutralino framework described in the previous section, 
we can estimate the number of observable neutralinos \chiobs{} 
for various dedicated LLP detectors at the LHC. We begin by calculating
the number of produced neutralinos via
\begin{equation}\label{eq:produced_neutralinos}
	N_{\neutralino{},\,\mathrm{M_{ab}},\,\mathrm{X}} =  N_{M_{ab}}\cdot \mathrm{Br}\big(M_{ab}\rightarrow \neutralino{}+ \mathrm{X}\big)\,,
\end{equation}
where $N_{M_{ab}}$ is the number of mesons $M_{ab}$ (see 
Table~\ref{tab:meson_numbers}) expected to be produced at the LHC 
during the runtime of the detector and $\mathrm{Br}\big(M_{ab}\rightarrow 
\neutralino{}+\mathrm{X}\big)$ is the branching ratio of a meson $M_{ab}$ 
to decay into a neutralino and other final state particles $\mathrm{X}$, 
\textit{cf.} Eq.~\eqref{eq:rpv-meson-decay}. If the neutralinos are 
sufficiently long-lived\footnotetext{Our approach goes beyond the implementation using \FORESEE{} in the simulation of proton-proton events. While the event generation is performed using \PYTHIA{}, we employ meson-production inputs consistent with those used in \FORESEE{}. In particular, the pion and kaon production numbers are based on cross sections obtained with \texttt{EPOS-LHC}.}, they travel a macroscopic distance and decay with
an average probability $\big<P\big(\neutralino{}\,\text{in f.v.} \big) 
\big>_{\mathrm{M_{ab}},\,\mathrm{X}}$ in the fiducial volume of the 
detector being considered. As we are specifically interested in the 
photon signature of the neutralino, \textit{cf.} 
Eq.~\eqref{eq:neutralinodecay}, we obtain \chiobs{} by
\begin{equation}\label{eq:observed_neutralinos}
	\chiobs{} = \mathrm{Br}\big(\neutralino{}\rightarrow \gamma +\overset{(-)}{\nu_i}\big) \cdot \sum_{M_{ab},\,\mathrm{X}}N_{\neutralino{},\,\mathrm{M_{ab}},\,\mathrm{X}}\cdot \big<P\big(\neutralino{}\,\text{in f.v.}\big)\big>_{\mathrm{M_{ab}},\,\mathrm{X}}\,.
\end{equation}
In order to estimate the average decay probability, we employ Monte-Carlo techniques. The probability of any neutralino to decay inside of a reference volume can be obtained using the exponential decay law
\begin{equation}\label{eq:individual_decay_probability}
	P_i\big(\neutralino{}\,\text{in f.v.}\big)=\exp\bigg( -\frac{L_{T,i}}{\lambda_i} \bigg)\bigg[1-\exp\bigg( -\frac{L_{I,i}}{\lambda_i} \bigg)\bigg]\,,
\end{equation}
where $L_{T,i}$ is the length the neutralino travels from its production 
vertex to the detector volume and $L_{I,i}$ is the length it would travel
inside the detector volume if it would not decay. $L_{I,i}$ takes 
into account the flight direction of the neutralino, as well as the 
extent of the detector being considered along that axis. $\lam_i$ is the
boosted decay length, which is determined by the Lorentz factor $\gamma
_i$, the speed $\beta_i$ and the total decay rate of the neutralino
\begin{equation}\label{eq:individual_boosted_decay_length}
	\lambda_i=\gamma_i\beta_i\,\tau_{\neutralino{}}=\frac{\gamma_i\beta_i}{\Gamma_{\mathrm{tot},\,\neutralino{}}}\,.
\end{equation}
Simulating $N^{\mathrm{MC}}_{M_{ab},\,\mathrm{X}}$ neutralinos for each production mode, the average decay probability is given by
\begin{equation}\label{eq:average_decay_probability}
	\big<P\big(\neutralino{}\,\text{in f.v.}\big)\big>_{\mathrm{M_{ab}},\,\mathrm{X}}=\frac{1}{N^{\mathrm{MC}}_{M_{ab},\,\mathrm{X}}}\sum_i^{N^{\mathrm{MC}}_{M_{ab},\,\mathrm{X}}}P_i\big(\neutralino{}\,\text{in f.v.}\big)\,.
\end{equation}
We note here, that the average decay probability has to be evaluated 
individually for each decay mode $M\rightarrow \neutralino{}+\mathrm{X}$, since 
the kinematic variables depend on the masses of the involved particles 
and  the number of particles involved. If we were to average only for the
\neutralino{} or average the probability for each initial meson $M$, we 
might overestimate $\big<P\big>$ by combining a large decay rate for one 
mode $\mathrm{X}_1$ with a high decay probability of another mode $\mathrm{X}_2$. 

We use the event generator \PYTHIA{}~\cite{Bierlich:2022pfr} to simulate 
$10^6$ proton-proton collisions with a center-of-mass energy of 
\SI{14}{\tera\eV}. Depending on the considered initial mesons $M_{ab}$, 
we activate the modules ``\texttt{SoftQCD:all}" 
($\pi$, $K$),\footnote{Pythia is generally tuned for the central region
of the LHC. The production of some particles such as $\pi^0$ is not 
well validated for $\eta>5$. We account for this by following the 
far-forward tuning described in Ref.~\cite{Fieg:2023kld} when 
investigating pions as initial mesons.}  
``\texttt{HardQCD:hardccbar}"\,($D$, $D_s$) or 
``\texttt{HardQCD:hardbbbar}"\,($B$). We let the mesons decay into a 
neutralino and some final state X. In order to improve statistics, we 
only allow for decays into a neutralino during the event generation and 
correct for the branching ratio after the simulation.

The lengths $L_{T,i}$ and $L_{I,i}$ are calculated employing the 
algorithm described in Ref.~\cite{Gunther:2023vmz}, using the 
\neutralino{} production vertex and the detector geometries. In addition 
to \FASER{} and \FASERTwo{} \cite{Feng:2017uoz,FASER:2018eoc}, for which 
the single photon signature was already investigated in 
Ref.~\cite{Dreiner:2022swd} using \FORESEE{} \cite{Kling:2021fwx} for 
event generation, here we also analyze the potential for this signature 
at \ANUBIS{}\,\cite{Bauer:2019vqk, Shah:2024fpl}, 
\CODEXB{}\,\cite{Gligorov:2017nwh, CODEX-b:2019jve, 
RodriguezFernandez:2024jcq}, \FACET{}\,\cite{Cerci:2021nlb}, \MAPP{}, 
\MAPPTwo{}\,\cite{Pinfold:2019zwp, MoEDAL-MAPP:2022kyr, 
Kalliokoski:2025ula} and \MATHUSLA{}\,\cite{Chou:2016lxi, Curtin:2018mvb,
MATHUSLA:2020uve,MATHUSLA:2025eth}. We simply compute event rates
and assume that these experiments are able to detect the photon plus 
missing energy signature of the neutralino decay. The implemented 
geometry for most detectors is described in 
Ref.~\cite{DeVries:2020jbs}.\footnote{The geometry of \ANUBIS{} is still
subject to change. Ref.~\cite{Bauer:2019vqk} proposed using the PX14 
access shaft above \ATLAS{} as a detection volume by lining the 
shaft with four \SI{1}{\meter} thick tracking stations at 
\SI{18.5}{\meter} intervals. Currently, the preferred configuration is 
lining the ceiling of the \ATLAS{} cavern UX15 with these tracking
stations to enhance the covered solid angle. As the ceiling configuration
is not yet finalised, we approximated the detection volume using a 
segment of a tube with an inner radius of \SI{20}{\meter}, a height of 
\SI{45}{\meter} and an azimuthal angle coverage of \SI{60}{\degree}.

For \MATHUSLA{}, 
the decay volume has been significantly reduced compared to what was 
assumed in Ref.~\cite{DeVries:2020jbs} (see Ref.~\cite{MATHUSLA:2025eth} 
for the current status), while the decay volume of \MAPPTwo{} has been 
extended to cover a greater part of the UGC1 gallery. \MAPP{} has been 
moved from the UGC1 gallery to the UA83 tunnel about \SI{100}{\meter} 
from IP8.}

An important difference between the event generation we employ here
compared to that used in Ref.~\cite{Dreiner:2022swd} is that \PYTHIA{} 
simulates actual decay vertices of the initial mesons. If $M_{ab}$ is 
long-lived, it decays a macroscopic distance away from the interaction 
point, which affects the distances $L_{T,i}$ and $L_{I,i}$. Additionally, 
we veto neutralino events depending on the macroscopic 
distance traveled by the meson, as described in 
Ref.~\cite{Gunther:2023vmz}. The displaced decay vertices together with 
the veto conditions affect the momentum spectrum of the neutralino we 
consider for the average decay probability and hence also \chiobs{}. We 
summarize the veto categories in Table~\ref{tab:detector_vetos}.
\begin{table}
	\centering
	\begin{tabular}{|c|cc|c|}
		\hline
		Category 												& Veto Conditions 				& Initial Meson & Applied to 		\\
		\hline
		\hline
		\multirow{2}{*}{Hadron Calorimetry (\ATLAS{})}	& $\rho\leq \SI{2.28}{\meter}$	& \multirow{2}{*}{All}			& \multirow{2}{*}{\ANUBIS{}}	\\
																& $z\leq \SI{4.277}{\meter}$	& 								& \\
		\hline
		\multirow{2}{*}{Hadron Calorimetry (\ATLAS{})}		& $\rho\leq \SI{1.77}{\meter}$	& \multirow{2}{*}{All}			& \multirow{2}{*}{\MATHUSLA{}}	\\
																& $z\leq \SI{4.0}{\meter}$		& 								& \\
		\hline
		TAN (IR1 / IR5)											& $z\leq \SI{141.2}{\meter}$	& Neutral Mesons				& \multirow{2}{*}{\FASER{}, \FACET{}}	\\
		TAS (IR1 / IR5)											& $z\leq \SI{19.05}{\meter}$	& Charged Mesons				& \\
		\hline
		\multirow{2}{*}{Shield Veto}							& $\rho\leq \SI{5.0}{\meter}$	& \multirow{2}{*}{All}			& \multirow{2}{*}{\CODEXB{}}	\\
																& $z\leq \SI{3.0}{\meter}$		& 								& \\
		\hline
		Cavern Wall												& $\rho\leq \SI{1.9}{\meter}$	& All							& \MAPP{}, \MAPPTwo{}	\\
		\hline
	\end{tabular}
	\caption{Summary of the vetos of Ref.~\cite{Gunther:2023vmz} implemented in the Monte Carlo simulation.}\label{tab:detector_vetos}
\end{table}

\section{Results}
\label{sec:results}

In this section, we present the results of our simulations for the six 
benchmarks proposed in Ref.~\cite{Dreiner:2022swd}. We summarize key 
characteristics of them in Table~\ref{tab:info_benchmarks}.
\begin{table}
	\centering
	\begin{tabular}{|c|cll|ccc|}
		\hline
		Benchmark	& $\lam_{P,iab}$			& \multicolumn{2}{c|}{$M_{ab}$}		& $\lam_{D,i'j'j'}$		& \multicolumn{2}{c|}{$\gamma$-signature} 		\\
		\hline
		\textbf{B1}	& $\lam_{211}^\prime$	& $\pi^\pm$,	& $\pi^0$		& $\lam_{333}^\prime$	& $\gamma+\nu_\mu$ ($d$),	& $\gamma+\nu_\tau$ ($b$)\\
		\textbf{B2}	& $\lam_{212}^\prime$	& $K^\pm$,		& $K_{L/S}$		& $\lam_{333}^\prime$	& \multicolumn{2}{c|}{$\gamma+\nu_\tau$ ($b$)}		\\
		\textbf{B3}	& $\lam_{112}^\prime$	& $K^\pm$,		& $K_{L/S}$		& $\lam_{322}$			& \multicolumn{2}{c|}{$\gamma+\nu_\tau$ ($\mu$)} 	\\
		\textbf{B4}	& $\lam_{221}^\prime$	& $D^\pm$,		& $K_{L/S}$		& $\lam_{233}$			& \multicolumn{2}{c|}{$\gamma+\nu_\mu$ ($\tau$)} 	\\
		\textbf{B5}	& $\lam_{222}^\prime$	& \multicolumn{2}{c|}{$D_s^\pm$}& $\lam_{222}^\prime$	& \multicolumn{2}{c|}{$\gamma+\nu_\mu$ ($s$)}		\\
		\textbf{B6}	& $\lam_{313}^\prime$	& $B^\pm$,		& $B^0$			& $\lam_{333}^\prime$	& \multicolumn{2}{c|}{$\gamma+\nu_\tau$ ($b$)}		\\
		\hline
	\end{tabular}
	\caption{Summary of the benchmarks proposed in Ref.~\cite{Dreiner:2022swd}. We list the non-zero couplings, the related mesons decaying into the neutralino, and the decays of neutralinos into the photon signature. In brackets, we signalize the fermion that is involved in the loop in the photon decay. See the respective subsections for more detail.}
	\label{tab:info_benchmarks}
\end{table}

\subsection{Benchmark 1}\label{sec:results:b1}

For the first benchmark we assume only $\lam^\prime_{211}$ and $\lam 
^\prime_{333}$ to be non-zero. The former coupling induces the decays 
$\pi^\pm\to\neutralino{}+\mu^\pm$ and $\pi^0\to\neutralino{}+\nu_\mu$, 
while both couplings allow for a radiative decay of the neutralino via 
either a $d$- or a $b$-quark loop. Since the 
radiative decay of the neutralino is proportional to $m_f^2$, where $m_f$ 
is the mass of the fermion in the loop, \textit{cf.} 
Eq.~\eqref{eq:neutralinodecay}, we allow for a non-zero $\lam^\prime_{333}$
in addition to $\lam^\prime_{211}$ to enhance the decay rate of the 
neutralino. We neglect contributions from $\rho$ ($\Upsilon$) mesons 
decaying into neutralinos via a non-zero $\lambda'_{211}$ ($\lambda'_{333}$). 
As discussed in  Ref.~\cite{deVries:2015mfw}, 
the lifetime of vector mesons 
is typically several orders of magnitude shorter than those of pseudoscalar 
mesons, so that the branching ratio of vector mesons into neutralinos 
and thus the production of neutralinos from vector mesons decays is 
negligible. Additionally, the production rate of $\Upsilon$ mesons at 
the LHC is significantly lower compared to lighter mesons, enhancing the suppression.
Hence, we omit these production channels from our simulation.

We are interested in a benchmark involving the production of the
light neutralinos via pions, since the latter are abundantly produced at 
the LHC. We note however, that the relatively large muon mass restricts
the kinematically available mass range for the neutralinos produced 
in $\pi^\pm$ decays, while the branching ratio of $\pi^0$ into neutralinos
is diminished due to the short lifetime of $\pi^0$. We could increase the 
neutralino mass range in charged decays by considering electrons and a 
non-zero $\lam^\prime_{111}$, but this coupling is heavily restricted from
neutrinoless double beta decay searches 
\cite{Hirsch:1995zi, Allanach:1999ic, Bolton:2021hje}.

\begin{figure}[ht]
	\centering
	\includegraphics[width=0.95\textwidth]{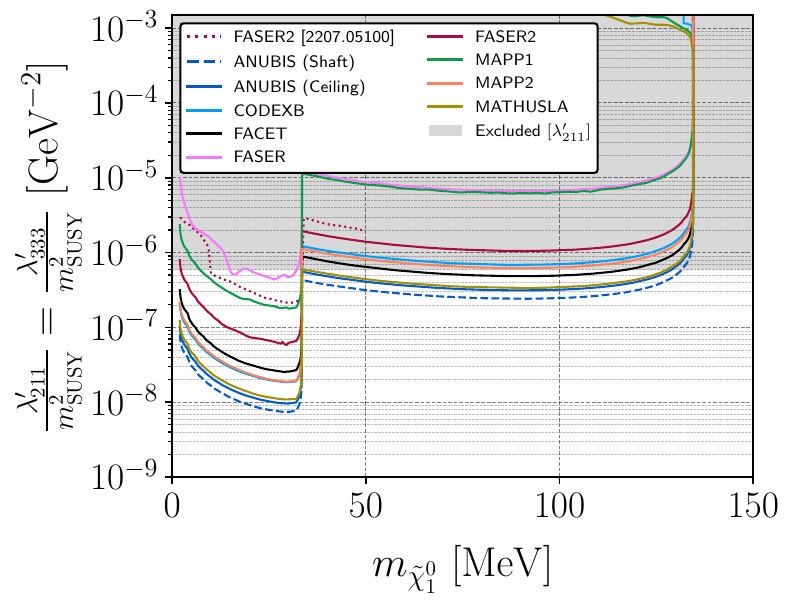}
	\caption{3-event isocurves in the $\lam/m_{\mathrm{SUSY}}^2$ vs. $m_{\neutralino{}}$ plane for benchmark 1. We assume mass degenerate sfermions and equal coupling strengths for $\lam^\prime_{211}$ and $\lam^\prime_{333}$. Previous results from Ref.~\cite{Dreiner:2022swd} are presented as dotted lines. The gray shaded area represents already excluded coupling regions, where the constraint for $\lam^\prime_{211}$ is the strongest constraint assuming $m_{\mathrm{SUSY}}=\SI{1}{\tera\eV}$.}\label{fig:benchmark_1}
\end{figure}

We present the results of the simulation in the $\lam'/m_{\mathrm{SUSY}} 
^2$ vs. $m_{\neutralino{}}$ plane in Fig.~\ref{fig:benchmark_1} as 
3-event isocurves, which correspond to a $95\%$ C.L.
exclusion limit assuming no observation and no background events. We
assume the coupling strengths of the non-zero couplings to be equal. The 
results of Ref.~\cite{Dreiner:2022swd} for \FASER{} and \FASERTwo{} are 
included as dotted lines and the already excluded 
region for the couplings is marked as 
a gray area. Assuming mass degenerate sfermions with 
$m_{\mathrm{SUSY}}=\SI{1}{\tera\eV}$, we have taken the 
stronger constraint listed in Table~\ref{tab:current_rpv_constraints} into
account.

In Fig.~\ref{fig:benchmark_1}, we can see that all detectors follow a 
similar pattern. Up to \linebreak \mbox{$m_{\neutralino{}}^{\mathrm{th}}\approx\SI{35}{\mega\eV}$}, the 
isocurves expand beyond already excluded areas, where the extent seems to 
favor larger neutralino masses. We can group the isocurves for all 
detectors as \ANUBIS{} and \MATHUSLA{} having the greatest sensitivity 
reach down  to $\lam^\prime/m^2\approx \SI{7.5e-9}{\per\giga\eV\squared}$, followed by \CODEXB{}, \FACET{} and 
\MAPPTwo{}. \FASERTwo{} can reach coupling strengths 
above $\SI{6e-8}{\per\giga\eV\squared}$, while
\MAPP{} and \FASER{} are barely sensitive below the existing coupling 
bound.\footnote{Some isocurves manifest fluctuations in the sensitivity 
reach, e.g. \FASER{}, \FASERTwo{} and \MAPP{}. These are not physical, but
rather due to limited statistics. These detector concepts are located 
furthest from the interaction vertex and cover only a small solid angle. 
Thus, the event rates are naturally small, requiring higher 
statistics for precise results.
A similar effect can be observed in the small decay length (large 
coupling) limit for all detectors. In this limit, neutralinos require a 
large boost to sufficiently contribute to $\big<P\big>$ and not decay 
before reaching the detector. Since the limit is already excluded by 
existing coupling constraints for all benchmarks, we do not address this 
issue by increasing the amount of simulated $pp$ collisions.}

The relative sensitivity reach between the detectors is determined by the 
considered integrated luminosity, the covered solid angle, the fiducial volume,
as well as the location of the detectors and the associated veto conditions. 
In this mass region, the main contribution to the observed events is from 
charged pion decays. These decays are prohibited for 
$m_{\neutralino{}}>m_{\neutralino{}}^{\mathrm{th}}$, 
so that only a couple of the detectors, namely \ANUBIS{}, \MATHUSLA{} and for 
some mass regions \FACET{}, can reach coupling regions beyond existing limits.

The sensitivity reach of \FASERTwo{} in our simulation is enhanced 
compared to previous results of Ref.~\cite{Dreiner:2022swd}. However, the 
dotted and solid isocurve are closer aligned for $m_{\neutralino{}}> m_ 
{\neutralino{}}^{\mathrm{th}}$, while they differ more significantly for 
$m_{\neutralino{}}<m_{\neutralino{}}^{\mathrm{th}}$. In 
Ref.~\cite{Dreiner:2022swd}, \FORESEE{} was used for event 
generation, where effects of long-lived initial mesons are taken into 
account by applying the exponential decay law to the decay probability. 
The exact decay position is not simulated and vetos based on the decay 
position were not included. Since here we do include the exact decay 
position, the detector sensitivity is affected in several ways, which
we now briefly discuss.
\begin{figure}
	\centering
	\includegraphics[width= 0.85\textwidth]{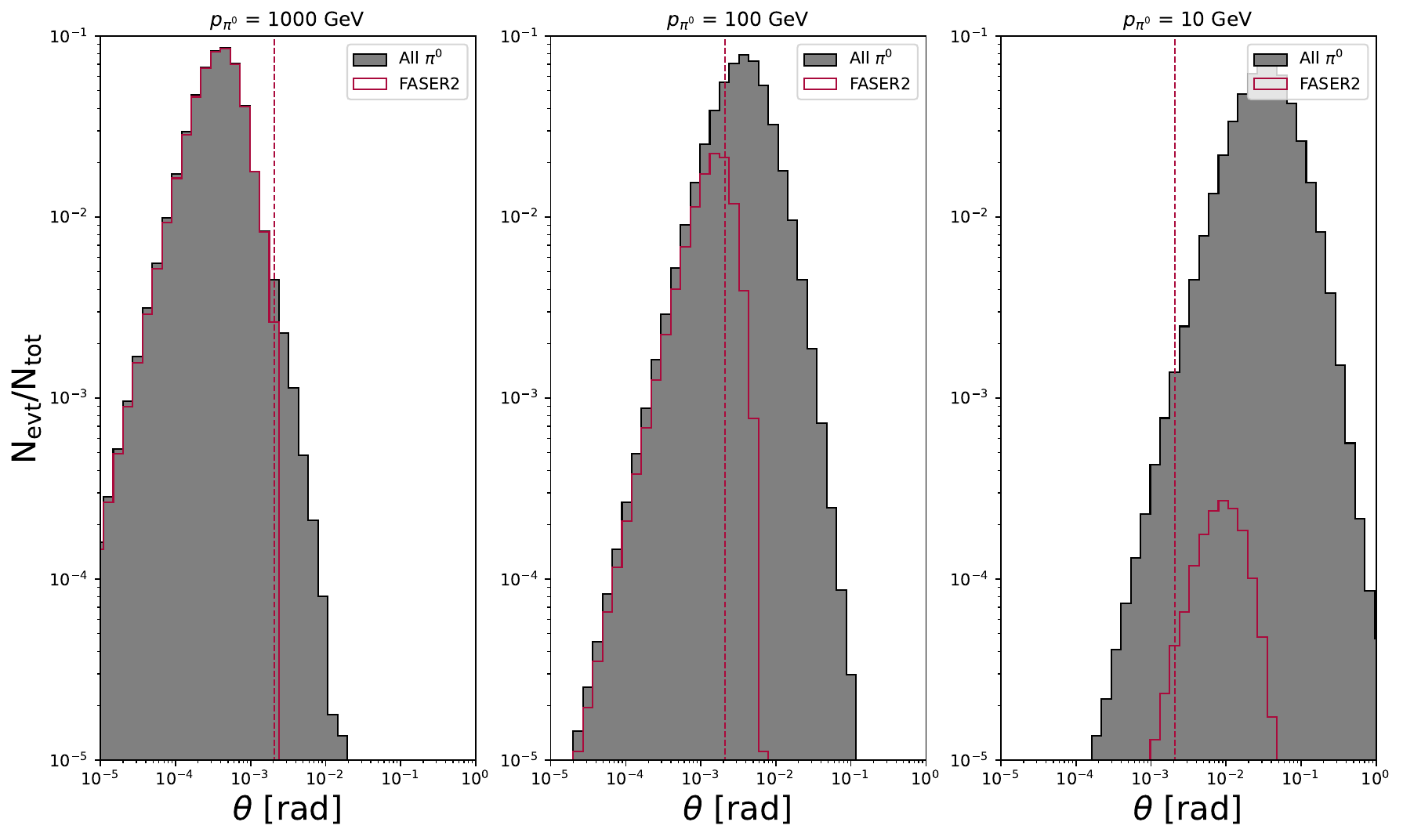}
	\caption{Normalized $\theta$-distributions of neutral pions generated in $N_{\mathrm{MC}}=10^6$ \SI{14}{\tera\eV} $pp$ events using the \texttt{SoftQCD} module of Pythia with the Forward Physics Tune \cite{Fieg:2023kld} for $p_{\pi^0}=$\SI{1000}{\giga\eV} (left), \SI{100}{\giga\eV} (middle) and \SI{10}{\giga\eV} (right). $\theta$ is the angle of the pion momentum with respect to the beam axis. The neutralino mass is set to $m_{\neutralino{}}=\SI{20}{\mega\eV}$. The gray histogram represents the $\theta$-distribution of all $\pi^0$, while the red histogram denotes the $\theta$-distribution within \FASERTwo{}. The red vertical dashed line depicts the angular acceptance of \FASERTwo{}.}
	\label{fig:normalized_pion_distribution}
\end{figure}

Firstly, mesons produced outside the geometric acceptance of 
\FASERTwo{} can nevertheless decay into neutralinos that propagate 
towards the detector. To illustrate this phenomenon, 
Fig.~\ref{fig:normalized_pion_distribution} depicts the normalized 
$\theta$-distributions of neutral pions. These have been generated in 
$N_{\mathrm{MC}}=10^6$ \SI{14}{\tera\eV} $pp$ events using the 
\texttt{SoftQCD} module of Pythia with the Forward Physics Tune~\cite{Fieg:2023kld}.
$\theta$ is the angle of the pion momenta with respect to the beam 
axis. For the momenta $p_{\pi^0}=\,$\SI{1000}{\giga\eV}, \SI{100} 
{\giga\eV} and \SI{10}{\giga\eV}, we show the $\theta$ distribution of
all pions (gray) together with that of pions decaying into neutralinos
that yield a non-zero contribution to $\big<P\big>$ for each detector 
(red). The geometric acceptance of \FASERTwo{} is included as 
a red dashed vertical line. Especially at lower momenta, mesons
outside the geometric acceptance contribute to the average decay 
probability. \FORESEE{} partially accounts for this effect by 
incorporating the kinematics of the neutralino, while assuming the 
production vertex coincides with the interaction point. We see 
that allowing for a production vertex displaced from the interaction 
point, for instance closer to the detector, can additionally enhance 
the effective geometric acceptance.
\begin{figure}
	\centering
	\includegraphics[width= 0.49\textwidth]{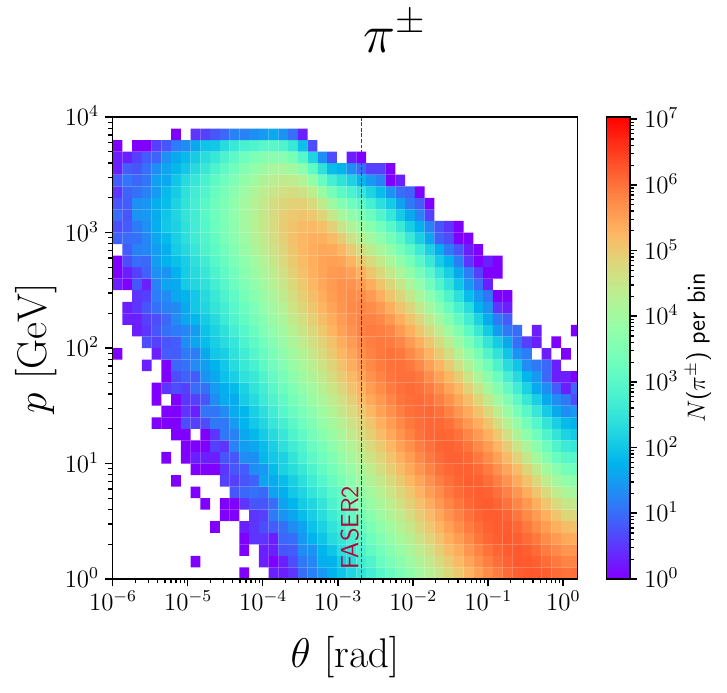}
	\includegraphics[width= 0.49\textwidth]{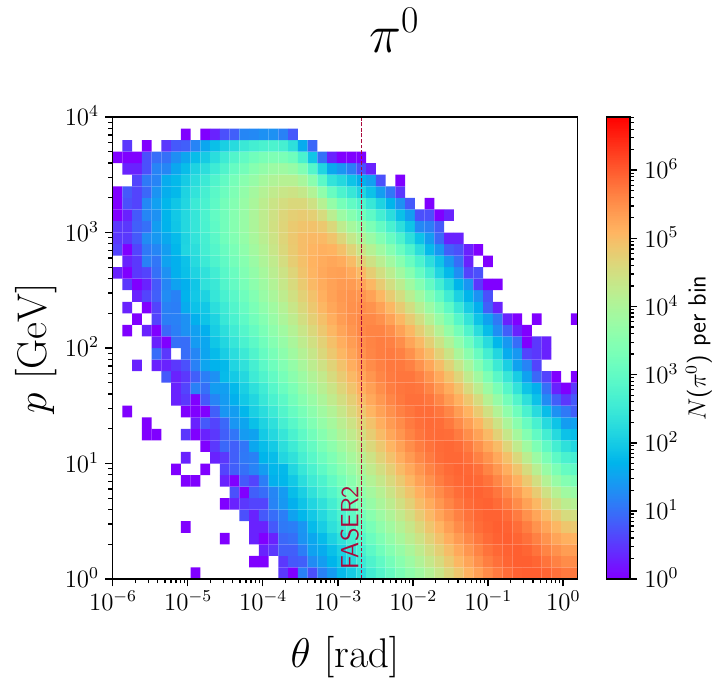}
	\caption{$(\theta,\,p)$-distribution of charged (left) and neutral (right) pions for $N_{\mathrm{MC}}=10^6$ \SI{14}{\tera\eV} $pp$ events which have been generated using the \texttt{SoftQCD} module of Pythia with the Forward Physics Tune \cite{Fieg:2023kld}. Here, $p$ is the absolute value of the momentum and $\theta$ is the angle of the pion momentum with respect to the beam axis. The red vertical dashed lines depict the angular acceptance of \FASERTwo{}.}\label{fig:pion_distribution}
\end{figure}

Secondly, neutralinos produced at larger 
$\theta$ exhibit a momentum spectrum shifted toward lower values than 
those originating in the far forward region (see 
Fig.~\ref{fig:pion_distribution} for the pion distributions). As 
demonstrated in Ref.~\cite{Beltran:2023nli}, in the large decay length
limit, \textit{i.e.} the small coupling limit, we can approximate the 
average decay probability from 
Eq.~\eqref{eq:average_decay_probability} to first order as
\begin{equation}\label{eq:probability_approximation}
	\big<P\big(\neutralino{}\,\text{in f.v.}\big)\big>_{\mathrm{M_{ab}},\,\mathrm{X}}\approx\frac{1}{N^{\mathrm{MC}}_{M_{ab},\,\mathrm{X}}}\sum_i^{N^{\mathrm{MC}}_{M_{ab},\,\mathrm{X}}}\frac{L_{I,i}}{\lam_i}=\frac{\Gamma_{\mathrm{tot},\,\neutralino{}}}{N^{\mathrm{MC}}_{M_{ab},\,\mathrm{X}}}\sum_i^{N^{\mathrm{MC}}_{M_{ab},\,\mathrm{X}}}\frac{L_{I,i}}{\gamma_i\beta_i}\,.
\end{equation}
We see, that a modified momentum spectrum alters $\big<P\big>$, as it 
is inversely proportional to the kinematic variables $\gamma_i$ and 
$\beta_i$. A lower average momentum increases the average decay 
probability in this limit. Hence, the sensitivity of \FASERTwo{}
is enhanced not only by the increased neutralino flux towards the 
detector, but also by the kinematic properties of the additional 
particles.

Lastly, $\big<P\big>$ is modified by the application of the 
veto conditions. For \FASERTwo{} for instance, the TAS (TAN) 
absorber, located \SI{19.05}{\meter} (\SI{141.2}{\meter}) 
downstream of the interaction point, restricts the allowed 
displaced decay vertices for charged (neutral) mesons. 
Neglecting the decay position relative to the surrounding 
infrastructure may therefore lead to an overestimate of the 
effective acceptance, particularly for highly boosted mesons 
decaying further away from the interaction point. In the
large decay length limit, mesons and subsequent neutralinos 
with large momenta contribute less to the average decay probability 
[see Eq.~\eqref{eq:probability_approximation}]. Overestimating the 
flux of high momenta neutralinos due to missing veto conditions does 
not compensate for the underestimation of low momenta neutralinos.

Thus, accounting for spatially displaced meson decay vertices at 
\FASERTwo{} enhances the detector sensitivity in scenarios with long-lived mesons producing neutralinos sufficiently displaced from the interaction point. Since $\tau_{\pi^\pm}\gg\tau_{\pi^0}$, charged pions typically decay farther from the interaction point, enhancing the impact of displaced meson decay vertices in $\pi^\pm$-dominated regions with $m_{\neutralino{}}<m_{\neutralino{}}^{\mathrm{th}}$.

The enhancing effect manifests itself further when comparing the 
relative ordering of \CODEXB{}, \FACET{}, and \MAPPTwo{} above and 
below $m_{\neutralino{}}^{\mathrm{th}}$ in Fig.~\ref{fig:benchmark_1}.
For $m_{\neutralino{}}>m_{\neutralino{}}^{\mathrm{th}}$, the largest 
sensitivity reach is achieved by \FACET{}, followed by \MAPPTwo{} and 
\CODEXB{}. However, for $m_{\neutralino{}}<m_{\neutralino{}} 
^{\mathrm{th}}$, where long-lived $\pi^\pm$ dominate neutralino 
production, \MAPPTwo{} and \CODEXB{} exhibit nearly identical 
sensitivities, whereas \FACET{} becomes less sensitive to RPV 
couplings. While \FACET{}, similar to \FASERTwo{}, is positioned in 
the far-forward region, \CODEXB{} is located at the largest angle with
respect to the beam axis of the three proposals. The sensitivity 
enhancement from displaced decays of long-lived $\pi^\pm$ increases 
with larger $\theta$ due to the shift of the meson momentum spectrum 
towards lower values. $\pi^0$ mesons are comparatively short-lived, 
leading to a lesser enhancement due to displaced decays. As a result, 
the difference between the regions below and above $m_{\neutralino{}} 
^{\mathrm{th}}$ is amplified at larger $\theta$ compared to the 
far-forward region, giving rise to the observed detector ordering.

\subsection{Benchmark 2}\label{sec:results:b2}

For the second benchmark, we investigate neutralinos produced by kaon 
decays instead of pions. Hence, we consider $\lam^\prime_{212}$ as the
non-zero production coupling, while $\lam^\prime_{333}$ induces the 
radiative decay. The neutralino is now produced via $K^\pm\to 
\neutralino{}+\mu^\pm$ and $K_{L/S}\to\neutralino{}+\nu_\mu$. The 
radiative decay via a $d$-quark loop is not possible in this 
benchmark.

\begin{figure}[ht]
	\centering
	\includegraphics[width=0.8\textwidth]{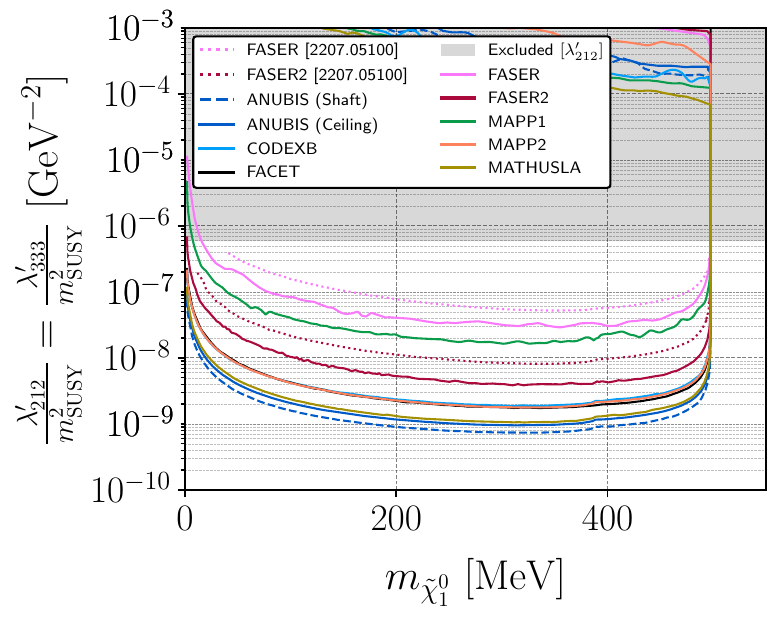}
	\caption{3-event isocurves in the $\lam/m_{\mathrm{SUSY}}^2$ vs. $m_{\neutralino{}}$ plane for benchmark 2. We assume mass degenerate sfermions and equal coupling strengths for $\lam^\prime_{212}$ and $\lam^\prime_{333}$. Previous results from Ref.~\cite{Dreiner:2022swd} are presented as dotted lines. The gray shaded area represents already excluded coupling regions, where the constraint for $\lam^\prime_{212}$ is the strongest constraint assuming $m_{\mathrm{SUSY}}=\SI{1}{\tera\eV}$.}\label{fig:benchmark_2}
\end{figure}

As for benchmark 1, we here present the results in 
Fig.~\ref{fig:benchmark_2} as 3-signal event isocurves 
assuming no background events. We include previous results as 
dotted lines and current coupling constraints as the gray 
area. The isocurves follow a similar structure for all 
detectors. The $m_{\neutralino{}}$ reach is bounded from above by 
the kaon mass. We observe a small kink at masses below 
\SI{400}{\mega\eV}, where the neutralino production via the 
decay of charged kaons is kinematically available. We 
furthermore observe a slight asymmetry around the point of 
maximum sensitivity, with an increased sensitivity towards 
higher masses $m_{\neutralino{}}$. The maximum sensitivity is 
observed for $m_{\neutralino{}} \approx \SI{350}{\mega\eV}$ with 
$\lam^\prime/m^2\approx \SI{7.5e-10}{\per\giga\eV\squared}$
for \ANUBIS{}. The various detectors can be grouped as in 
benchmark 1, where \ANUBIS{} and \MATHUSLA{} show the greatest
sensitivity, followed by \CODEXB{}, \FACET{} and \MAPPTwo{}, 
then \FASERTwo{}, and finally \MAPP{} and \FASER{} show the 
lowest sensitivity. 

We can also observe the effects of displaced kaon decays when comparing the results of our simulation with the previous investigation using \FORESEE{}. As $K^\pm$ and $K_L$ are long-lived, the difference between both investigation can be explained as in benchmark 1. The neutralino momentum spectrum and the neutralino flux are modified by considering the explicit kaon decay vertices and the associated veto conditions in a way to enhance $\big<P\big>$.

\subsection{Benchmark 3}\label{sec:results:b3}

In this benchmark we select $\lam^\prime_{112}\not=0$ for the 
production of the neutralinos via kaon decays:
\begin{equation}
    K^\pm\to \neutralino{} + e^\pm\,,\quad K^0\to \neutralino{} +\nu_e\,.
\end{equation}
For the neutralino decay we consider $\lam_{322}\neq 0$. Leading to 
the radiative and tree-level decays
\begin{equation}
    \neutralino{}\to \left\{
    \begin{array}{l}
         \nu_\tau +\gamma\,,\;\text{via a  ($\mu-\tilde{\mu}$)-loop}\,, \\
         \nu_\tau + \mu^\pm + \mu^\mp\,. 
    \end{array}
    \right.
\end{equation}
For production via kaons the tree-level decay $\neutralino{}\to\nu_\mu
+\tau^\pm + \mu^\mp$ is kinematically not possible. The additional 
neutralino decays reduce the branching fraction for the radiative 
decays.

\begin{figure}[ht]
	\centering
	\includegraphics[width=0.8\textwidth]{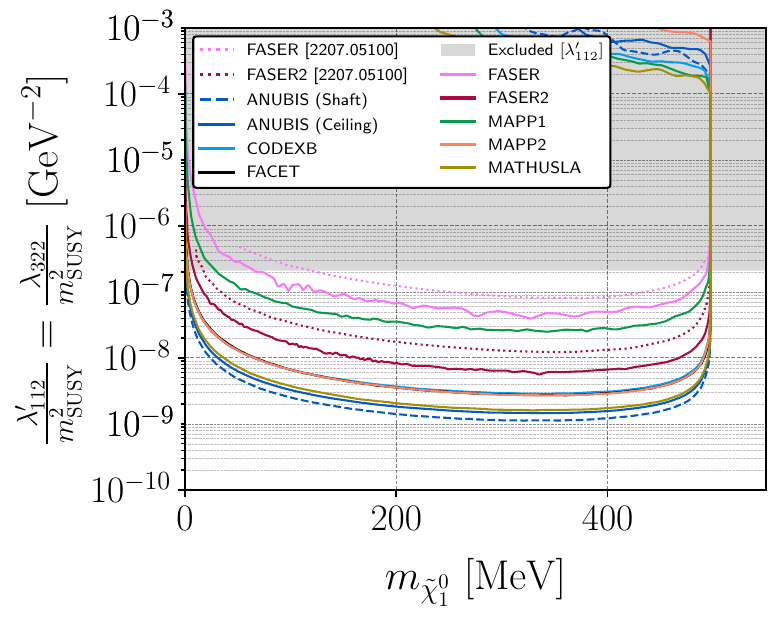}
	\caption{3-event isocurves in the $\lam^{(\prime)}/m_{\mathrm 
	{SUSY}}^2$ vs. $m_{\neutralino{}}$ plane for benchmark 3. We 
	assume mass degenerate sfermions and equal coupling strengths for $\lam^\prime_{112}$ and $\lam_{322}$. Previous results from Ref.~\cite{Dreiner:2022swd} for \FASER{}/\FASERTwo{} are 
	presented as dotted lines. The gray shaded area represents already excluded coupling regions, where the constraint for $\lam^\prime_{112}$ is the strongest 
	constraint assuming $m_{\mathrm{SUSY}}=\SI{1}{\tera\eV}$.}\label{fig:benchmark_3}
\end{figure}

Results for benchmark~3 (Fig.~\ref{fig:benchmark_3}) are very similar 
to benchmark~2 although overall the sensitivity is lower. The kink for
masses below \SI{400}{\mega\eV} can no longer be observed, as the 
charged decays are already kinematically available, but we still 
observe the enhanced sensitivity for large $m_{\neutralino{}}$. The 
most sensitive detector is \ANUBIS{} with a reach of $\lam^\prime 
/m^2\approx \SI{e-9}{\per\giga\eV\squared}$, while 
\MAPP{} and \FASER{} are the least sensitive. We can also observe the 
effects of the displaced decays, when comparing to previous results.

\subsection{Benchmark 4}\label{sec:results:b4}

To investigate heavier neutralinos, we now consider charm
meson decays with a non-zero $\lam^\prime_{221}$. 
This coupling allows for the production of the neutralinos via
the decay $D^\pm\to \neutralino{}+\mu^\pm$, as well as 
via the neutral kaon decays $K_{L/S}\to\neutralino{} 
+\nu_\mu$. This implies $m_{\neutralino{}} < m_{\tau}$. If the neutralino is heavy enough the latter 
production mode becomes kinematically inaccessible, but neutralino
decays into kaons such as $\neutralino{}\to K_{L/S}+\nu_\mu$ 
and $\neutralino{}\to K^{*0} + \nu_\mu$ become allowed. For $m_{\neutralino{}}>\orderof(\SI{1.5}{\giga\eV})$, 
decays involving multi-meson final states might become 
relevant, but are neglected here for simplicity. As can 
be seen in Eq.~\eqref{eq:probability_approximation}, a larger 
total decay rate can increase the average decay probability, while it would reduce the 
branching ratio into the signature single-$\gamma$ 
decay. Quantifying this effect requires a detailed analysis, which goes beyond 
the scope of the present work. For the radiative decay, we again consider an 
$LL\bar{E}$ coupling, $\lam_{233}$, where the loop is mediated
by a $\tau-\tilde\tau$ pair. Fully leptonic decays are 
not possible because they always include at least one $\tau$ 
and $m_{\neutralino{}}<m_\tau$ for all available neutralino 
production modes. 

\begin{figure}[ht]
	\centering
	\includegraphics[width=0.8\textwidth]{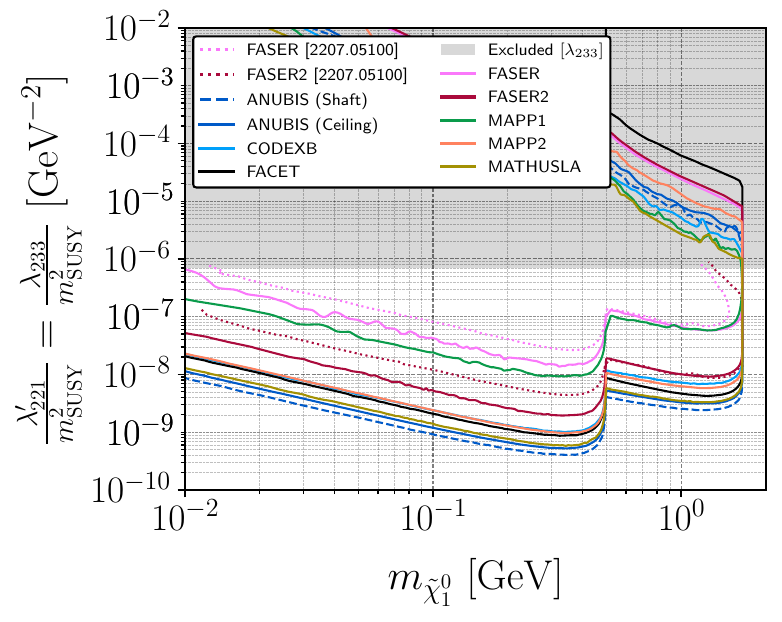}
	\caption{3-event isocurves in the $\lam/m_{\mathrm{SUSY}}^2$ vs. $m_{\neutralino{}}$ plane for benchmark 4. We assume mass degenerate sfermions and equal coupling strengths for $\lam^\prime_{221}$ and $\lam_{233}$. Previous results from Ref.~\cite{Dreiner:2022swd} are presented as dotted lines. The gray shaded area represents already excluded coupling regions, where the constraint for $\lam_{233}$ is the strongest constraint assuming $m_{\mathrm{SUSY}}=\SI{1}{\tera\eV}$.}\label{fig:benchmark_4}
\end{figure}

We have chosen to depict the results of benchmark 4 
(Fig.~\ref{fig:benchmark_4}) as a function of the logarithm of $m_ {\neutralino{} 
}$. We can divide the sensitivity reach into $m_{\neutralino{}} 
$-regions where the production is dominated by kaon 
decays and $m_{\neutralino{}}$-regions where it can 
kinematically only proceed via the decay of charm mesons. As 
charm mesons are less abundantly produced at the LHC,
the detectors are less sensitive to RPV couplings in the 
latter case. As expected from the previous benchmarks, 
\ANUBIS{} is the most sensitive detector concept reaching 
couplings as low as $\lam^\prime/m^2\approx\SI{4e-10}{\per 
\giga\eV\squared}$, while the other detector concepts align
as in previous benchmarks.

Comparing the isocurves for \FASER{} and \FASERTwo{} with 
previous investigation, we see an enhanced 
sensitivity in our results for $m_{\neutralino{}}<\SI{500} 
{\mega\eV}$. However, the curves agree for 
heavier neutralinos. Recall, we have an improved 
simulation procedure. Charm mesons decay promptly, so that 
the meson decay vertices are almost identical to the 
interaction point when considering macroscopic scales. Hence, 
the considered neutralino spectrum and our event generation is
similar to Ref.~\cite{Dreiner:2022swd}.

\subsection{Benchmark 5}\label{sec:results:b5}

In benchmark five, we only consider a single non-zero coupling
$\lam^\prime_{222}$. This coupling induces the production
of the neutralinos via the decay $D_s^\pm\to\neutralino{}+\mu 
^\pm$ as well as the decay via the radiative mode 
$\neutralino{}\to  \gamma + \nu_\mu$ via an s-quark loop. 
Additionally, the neutralino can decay into $\eta$, $\eta^ 
\prime$, or $\phi$ for a large enough $m_{\neutralino{}}$.

\begin{figure}[ht]
	\centering
	\includegraphics[width=0.8\textwidth]{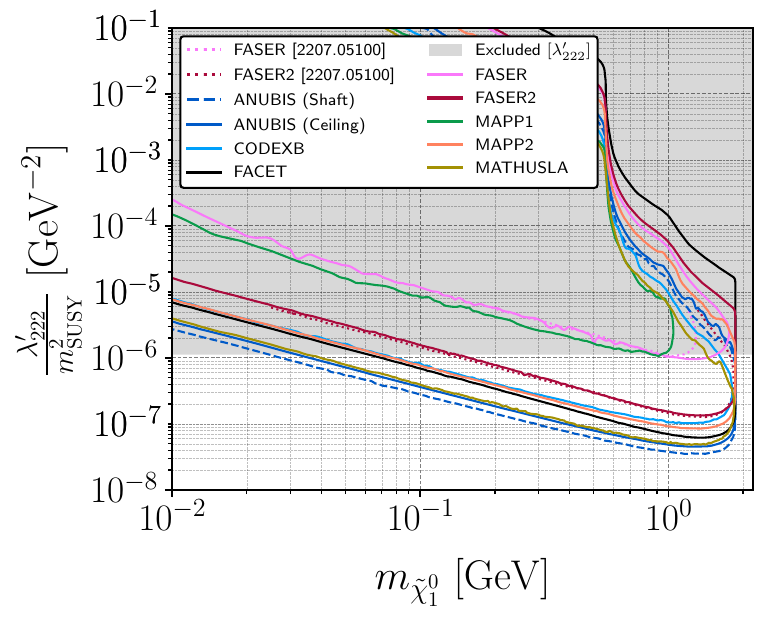}
	\caption{3-event isocurves in the $\lam/m_{\mathrm{SUSY}}^2$ vs. $m_{\neutralino{}}$ plane for benchmark 5. We assume mass degenerate sfermions. Previous results from Ref.~\cite{Dreiner:2022swd} are presented as dotted lines. The gray shaded area represents already excluded coupling regions for $\lam^\prime_{222}$ assuming $m_{\mathrm{SUSY}}=\SI{1}{\tera\eV}$.}\label{fig:benchmark_5}
\end{figure}

Fig.~\ref{fig:benchmark_5} displays the resulting 3-event 
isocurves. We can observe various features also 
present in the other benchmarks such as the order of sensitivity 
between the detectors. \ANUBIS{} has the greatest sensitivity 
reach with $\lam^\prime/m^2\approx\SI{3.5e-8}{\per\giga \eV 
\squared}$, whereas \MAPP{} and \FASERTwo{} are only sensitive
up to existing limits. Since the $D_s^\pm$ decay promptly, we 
do not observe the effects of displaced meson decays on the 
neutralino spectrum, so that our simulation results agree with
Ref.~\cite{Dreiner:2022swd}. Decays of the neutralinos into 
$\eta$, $\eta^\prime$, or $\phi$ do not affect the shape of 
the isocurves for small couplings, while they influence the small decay 
length limit. Mass thresholds of these mesons can be observed 
in the already excluded regions.

\subsection{Benchmark 6}\label{sec:results:b6}

Lastly, we allow for B-mesons to decay into the neutralino via
$B^\pm\to\neutralino{}+\tau^\pm$ and $B^0\to\neutralino{}+\nu 
_\tau$ by setting $\lam^\prime_{313}\neq 0$. The only 
kinematically allowed decay mode is the radiative mode 
via the b-quark loop induced by $\lam^\prime_{333}$.

\begin{figure}[ht]
	\centering
	\includegraphics[width=0.8\textwidth]{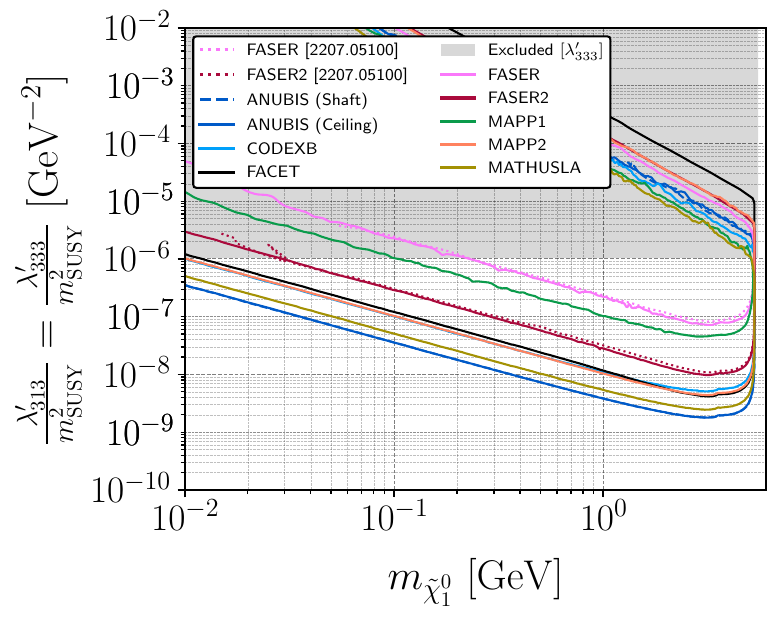}
	\caption{3-event isocurves in the $\lam/m_{\mathrm{SUSY}}^2$ vs. $m_{\neutralino{}}$ plane for benchmark 6. We assume mass degenerate sfermions and equal coupling strengths for $\lam^\prime_{313}$ and $\lam^\prime_{333}$. Previous results from Ref.~\cite{Dreiner:2022swd} are presented as dotted lines. The gray shaded area represents already excluded coupling regions, where the constraint for $\lam^\prime_{333}$ is the strongest constraint assuming $m_{\mathrm{SUSY}}=\SI{1}{\tera\eV}$.}\label{fig:benchmark_6}
\end{figure}

Isocurves for this benchmark (Fig.~\ref{fig:benchmark_6}) 
depict a lot of characteristics we discussed in the previous 
benchmarks. Because of the large tau mass, we observe a small kink at the kinematic threshold for the 
charged decays. The greatest sensitivity reach is observed for
\ANUBIS{} with $\lam^\prime/m^2\approx\SI{1.9e-9}{\per\giga\eV 
\squared}$, while \FASER{} is the least sensitive. As in benchmark 4, for 
$m_{\neutralino{}}>\SI{500}{\mega\eV}$ and benchmark 5, the 
initial mesons decay promptly, so that displaced decays do not
affect the isocurves. Hence, our simulation agrees with the 
results for \FASER{} and \FASERTwo{} of 
Ref.~\cite{Dreiner:2022swd}.

\section{Conclusions}
\label{sec:conclusion}
We have analyzed the case of a light [$m_{\neutralino{}} 
<\mathcal{O}(1.5\,\mathrm{GeV})$] long-lived neutralino 
within supersymmetry with broken R-parity in the context of 
the LHC at CERN, with additional remote detectors:  \ANUBIS{},
\CODEXB{}, \FACET{}, \FASER{}, \FASERTwo{}, \MAPP{}, 
\MAPPTwo{}, and \MATHUSLA{}. Such a light neutralino is 
consistent with all current experimental constraints. We have
simulated the production of the neutralinos via rare meson 
decays and defined six benchmark scenarios according to the
dominant production channel: pions, kaons, a mixture of kaons 
and $D$-mesons, just $D$-mesons, or via $B$-mesons. In all 
cases, we have chosen the solely non-vanishing R-parity 
violating coupling such that the radiative neutralino 
decay $\neutralino{}\to \gamma+\nu$ is dominant. The 
corresponding benchmark characteristics are summarized in 
Table~\ref{tab:info_benchmarks}. The main purpose of this 
work is to analyze the sensitivity of the above detectors to 
this radiative mode. We have extended the previous work in 
Ref.~\cite{Dreiner:2022swd}, which primarily focused on
\FASER{} and \FASERTwo{}, to analyze the other proposed 
detectors. In addition, we have improved the simulation by 
taking into account the finite decay length of the 
parent  meson, which enhances the sensitivity as discussed in Sec.~\ref{sec:results:b1} and depicted in Figs.~\ref{fig:normalized_pion_distribution} and 
\ref{fig:pion_distribution}.

Our main results are presented in Fig.~\ref{fig:benchmark_1} 
for benchmark 1, Fig.~\ref{fig:benchmark_2}, for benchmark 2,
Fig.~\ref{fig:benchmark_3}  for benchmark 3,
Fig.~\ref{fig:benchmark_4} for benchmark 4, Fig.~\ref{fig:benchmark_5} for benchmark 5, and
Fig.~\ref{fig:benchmark_6} for benchmark 6. Overall, accounting
for existing low-energy bounds on the R-parity violating 
couplings, we find for all benchmarks and for all detectors an 
unexplored sensitivity range, allowing for potential 
discoveries. Among the considered experiments, \ANUBIS{} 
generally is the most sensitive and \FASER{} the least 
sensitive experiment to the scenarios studied here, while of course \FASER{} has already taken data.

\section*{Acknowledgments}
HKD would like to thank the Universiteit van Amsterdam and the 
Nikhef Theory Group for their hospitality while part of this 
work was completed. We thank Felix Kling, Florian 
Bernlochner, and Matthias Schott for discussions.

\providecommand{\href}[2]{#2}\begingroup\raggedright\endgroup

\end{document}